\documentclass{osa-article}
\journal{oe}

\usepackage[utf8]{inputenc}
\usepackage[T1]{fontenc}

\usepackage{amsmath,amssymb,amsfonts}
\usepackage{amstext, mathrsfs, textcomp}
\usepackage{mathptmx}
\usepackage{bigints}  
\usepackage{nicefrac}
\usepackage{multirow}
\usepackage{dcolumn}

\usepackage{bm}

\usepackage{graphicx}
\usepackage{xcolor}
\usepackage{hyperref}

\usepackage[final]{changes}

\graphicspath{{Figures/}{TOC_ACS/}}

\newcommand{\FIGCAPTIONPREFIX}{}

\newcommand{\TITLE}{Design of plasmonic directional antennas via evolutionary optimization}

\newcommand{\onlinecite}{\cite}

\begin{document}
\title{\TITLE}

\author{Peter R. Wiecha$^{1,2,*}$, Clément Majorel$^2$, Christian Girard$^2$, Aurélien Cuche$^2$, Vincent Paillard$^2$, Otto L. Muskens$^1$, and Arnaud Arbouet$^2$}

\address{$^1$Physics and Astronomy, Faculty of Engineering and Physical Sciences, University of Southampton, Southampton, UK\\
$^2$CEMES-CNRS, Universit\'e de Toulouse, CNRS, UPS, Toulouse, France\\}

\email{$^*$p.wiecha@soton.ac.uk}



\begin{abstract}
We demonstrate inverse design of plasmonic nanoantennas for directional light scattering. 
Our method is based on a combination of full-field electrodynamical simulations via the Green dyadic method and evolutionary optimization (EO).
Without any initial bias, we find that the geometries reproducibly found by EO, work on the same principles as radio-frequency antennas. 
We demonstrate the versatility of our approach by designing various directional optical antennas for different scattering problems.
EO based nanoantenna design has tremendous potential for a multitude of applications like nano-scale information routing and processing or single-molecule spectroscopy.
Furthermore, EO can help to derive general design rules and to identify inherent physical limitations for photonic nanoparticles and metasurfaces.
\end{abstract}


The vast opportunities of light-based nanotechnology lead to tremendous research efforts on light-matter interaction at sub-wavelength scales, and in particular gave rise to the broad field of plasmonics, as localized surface plasmon-polariton (LSP) resonances in metallic nanostructures can strongly confine far-field radiation \cite{muhlschlegel_resonant_2005}.
By variations of the particle geometry or its material it is possible to tailor manifold optical properties like resonance positions \cite{tan_plasmonic_2014}, polarization conversion \cite{wiecha_polarization_2017}, optical chirality \cite{valev_chirality_2013}, localized heat generation \cite{baffou_heat_2009, girard_designing_2018} or nonlinear optical effects \cite{kauranen_nonlinear_2012, mesch_nonlinear_2016} with many applications like field enhanced spectroscopy \cite{vercruysse_directional_2014} or refractive index sensing \cite{wersall_directional_2014, mesch_nonlinear_2016}.

Another optical functionality that is in the focus of the plasmonics community is the directional routing of light.
Not only individual nanostructures were demonstrated for directional scattering of far-field light \cite{vercruysse_unidirectional_2013, abass_insights_2016, yao_controlling_2016}, color-routing \cite{shegai_bimetallic_2011}, quantum emitter radiation steering \cite{curto_unidirectional_2010, maksymov_optical_2012, vercruysse_directional_2014, hancu_multipolar_2014, dregely_imaging_2014, ramezani_hybrid_2015}, electro-luminescence \cite{gurunarayanan_electrically_2017, r.__kullock_directed_2018} or directional non-linear emission \cite{xiong_compact_2016}.
Also metasurfaces for directional scattering and color-routing have been proposed \cite{lindfors_imaging_2016, yan_fano-resonance-assisted_2017}.
Such photonic devices imply applications like on-chip light routing and de-multiplexing \cite{guo_highbit_2017} or sensing \cite{vercruysse_directional_2014, wersall_directional_2014}.

\begin{figure}[tb]
	\centering
	\includegraphics{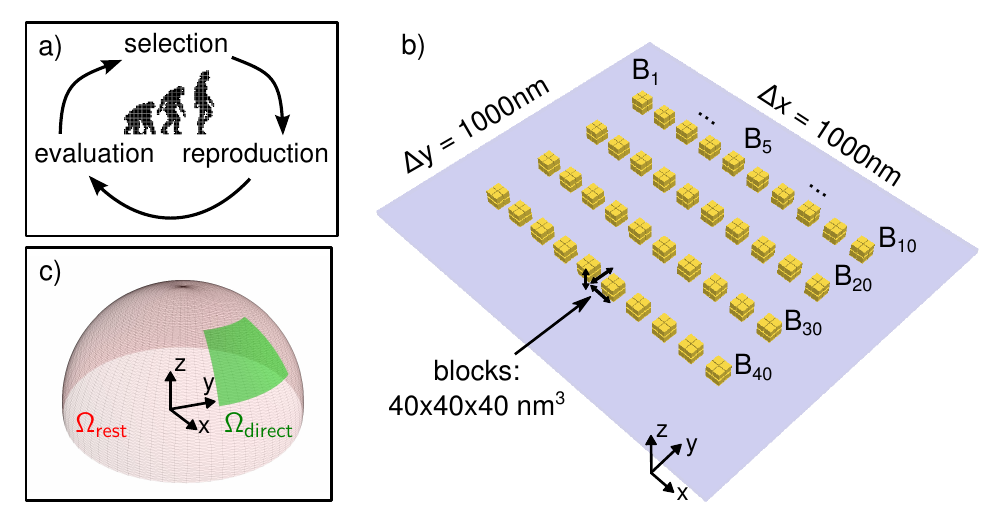} 
	\caption{\FIGCAPTIONPREFIX
		(a) Scheme illustrating evolutionary optimization.
		(b) Plasmonic antenna structure model for EO of directional scattering. \(40\) gold-blocks (B\(_i\)), each \(40\times40\times40\,\)nm\(^3\) large are placed on a \(n_{\text{s}}=1.5\) substrate (in the \(XY\) plane) within an area of \(1 \times 1\,\)\textmu m\(^2\). Plane wave illumination at normal incidence (\(\mathbf{k}\) along \(-z\)), with \(\lambda_0=800\,\)nm, linearly polarized along \(OX\).
		(c) Sketch of the directionality problem: Maximize the ratio of scattered intensity through a narrow window (green) and scattering to the remaining solid angle (red).
	}
	\label{fig:fig1}
\end{figure}

To obtain functionalities such as directional scattering of light, a careful design of the nano-antenna is crucial. 
In the conventional approach a reference geometry is chosen from qualitative considerations whose properties are subsequently studied systematically. However, imposing a geometry from the start risks excluding interesting solutions to the problem.
A promising alternative can be evolutionary optimization (EO) strategies. 
By mimicking natural selection through a cycle of reproduction, evaluation and selection (see Fig.~\ref{fig:fig1}(a)), EO is able to find global optima for complex non-analytical problems \cite{sivanandam_introduction_2008}.
EO has recently been applied very successfully to various problems in nano-optics like maximization of local field enhancement \cite{forestiere_particle-swarm_2010, feichtner_evolutionary_2012, forestiere_inverse_2016} or of optical resonances and structural color \cite{ginzburg_resonances_2011, bigourdan_design_2014, mirzaei_superscattering_2014, wiecha_evolutionary_2017, girard_designing_2018, gonzalez-alcalde_optimization_2018}.
Other applications include the design of carbon nanotube field-emission sources \cite{chen_optimal_2007}, dielectric anti-reflection coatings \cite{good_general_2016}, optical cloaking \cite{mirzaei_all-dielectric_2015}, multi focal-spot flat lenses \cite{hu_evolutionary_2017} or photonic power-splitters \cite{molesky_inverse_2018}.

In this paper we use EO to design directional optical antennas. 
We couple an evolutionary algorithm to full-field electro-dynamical simulations using the Green dyadic method (GDM).
With a randomly initialized and almost unconstrained geometric model, our approach reproducibly yields antennas resembling the classical radio-frequency (RF) Yagi-Uda layout \cite{yagi_beam_1928}.
We find, that the working principle of these plasmonic antennas is indeed \added{equivalent} to their RF counter-parts.
We demonstrate the versatility of our method on different directional scattering problems such as varying angles of scattering, double-focusing, or quantum emitter steering. 
We finally show that evolutionary optimization can yield general design rules and furthermore can be used to identify and analyze physical limitations in nanoantenna design.

\begin{figure}[tb]
	\centering
	\includegraphics[width=\textwidth]{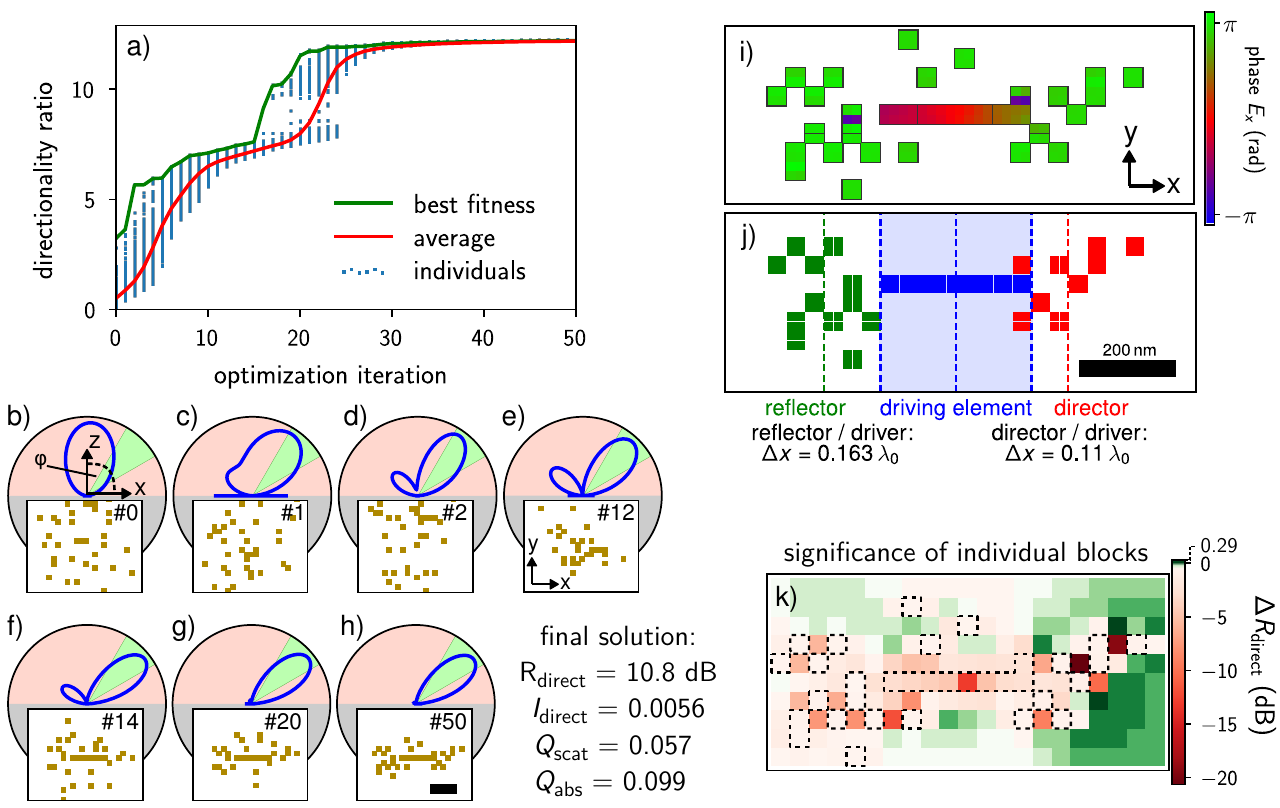} 
	\caption{\FIGCAPTIONPREFIX
		(a) Fitness vs. EO iteration with a population of \(N_{\text{pop}}=500\) individuals. 
		Best and average fitness are indicated by a green, respectively red line.
		(b-h) Best solutions at selected iterations between random initialization (b) and final solution (h). Iteration numbers are indicated at the bottom left. Scattering patterns in \(XZ\)-plane are shown in the top panels by blue lines, a green segment indicates the optimization target. The corresponding gold cube arrangements are shown in the panels below.
		(i) Phase of the electric field \(X\) component relative to the driving element's center. 
		(j) Functional components: feed (blue), reflector (green) and director (red). 
		The distances between the centers of gravity (dashed lines) are indicated at the top. 
		(k) Impact on the directionality ratio in decibel ($10 \log_{10} (R_{\text{direct}})$) when each block is toggled (gold block \(\leftrightarrow\) no gold block).
		Scale bars are \(200\,\)nm.
	}
	\label{fig:fig2}
\end{figure}

\section{Optimization method, model and problem}

\added{Evolutionary optimization is a heuristic approach to search for an optimum solution of a complex problem, which cannot be optimized by classical means such as steepest descent methods. 
EO considers a certain number of randomly generated parametersets describing a problem. Parameters may be for instance the positions or sizes of sub-constituents of a nanostructure.
The ensemble of parametersets is called the \textit{population}.
In a first step, each parameterset (also called an \textit{individual}) is evaluated, in our case through a numerical simulation of the optical scattering (more details below).
The evaluation criterion is in general referred to as the \textit{fitness function}.
In a \textit{selection} step, the best solutions of the population (\textit{i.e.} with the highest fitness values) are chosen for the following \textit{reproduction} step.
Therein, randomly chosen parameters are mutated (randomly modified) and interchanged between individuals, mimicking evolutionary processes in nature and yielding a new population of parametersets.
This cyclic process is repeated multiple times (see Fig.~\ref{fig:fig1}(a)) until some stop criterion like a timelimit is met, leading to a close-to-optimum solution to the problem.}

As EO algorithm we use the implementation of self-adaptive differential evolution (``jDE'') in the ``pyGMO'' toolkit \cite{biscani_global_2010}.
It takes as input the directionality of scattering, calculated using ``pyGDM'' \cite{wiecha_pygdmpython_2018}, our own python implementation of the Green Dyadic Method (GDM) \cite{girard_shaping_2008}. 
By including a surface propagator we can take into account a glass substrate \cite{girard_generation_1995}.
Details on our EO-GDM technique can be found elsewhere \cite{wiecha_evolutionary_2017, wiecha_pygdmpython_2018}.
\added{We want to note, that any numerical approach for solving Maxwell's equations can be used together with EO (possible other methods are for example finite difference time domain \cite{feichtner_evolutionary_2012} or mode expansion techniques \cite{mirzaei_superscattering_2014}).
With the GDM, we use a volume integral method which is particularly convenient when a relatively small nanostructure is placed in a large domain, where the GDM provides fast convergence.}

A scheme of the geometric model to be optimized is shown in Fig.~\ref{fig:fig1}(b).
\(40\) gold blocks B\(_i\), each \(40 \times 40 \times 40\,\)nm\(^3\) in size, are placed in air (\(n_{\text{env}}=1\)) on a glass substrate (\(n_{\text{s}} = 1.5\)). For gold, we use the tabulated dispersion from Ref.~\onlinecite{johnson_optical_1972}.
Each of the blocks are modeled by \(2 \times 2 \times 2\) dipoles. 
\added{A discussion on the discretization step and the elementary block size and shape can be found in the appendix (Fig.~\ref{figApp:discretization_step}).}
The optimization target parameters are the positions \((x_i, y_i)\) of the blocks, which are bound to an area of \(1 \times 1\,\)\textmu m\(^2\).
Additionally, the positions of the blocks are constrained to lie on a grid with steps of the block-size (\(40\,\)nm), to avoid the problem of partially overlapping elements. 
This results in a total of \(25\times 25\) possible positions.
We note that blocks with identical positions are treated as a single block, hence the amount of material is not strictly constant.
With \(40\) gold blocks, we have~\(80\) free parameters and \((25\times 25)! / (25\times 25 - 40)! \approx 10^{111}\) possible arrangements.

A plane wave with \(\lambda_0 = 800\,\)nm, linearly polarized along \(OX\), is normally incident onto the gold blocks. 
Using the GDM, we calculate at first the average intensity per solid angle \(I_{\text{direct}}\), scattered through a solid angle \(\Omega_{\text{direct}}\) of polar and azimuthal dimensions \(\Delta\varphi = 30\,^{\circ}\) and \(\Delta\theta = 45\,^{\circ}\), centered at \(\varphi = 45\,^{\circ}\) and \(\theta = 0\,^{\circ}\) (diagonal between \(X\) and \(Z\) axis), as illustrated by the green window in Fig.~\ref{fig:fig1}(c).
Second we calculate \(I_{\text{rest}}\), the average intensity per solid angle, scattered towards the remaining hemisphere (\(\Omega_{\text{rest}}\), red in Fig.~\ref{fig:fig1}(c)).
The goal of the EO is to maximize the ratio between scattering per solid angle to the target ``window'' and to the rest of the hemisphere:
\begin{equation}
  R_{\text{direct}} = \dfrac{\Big( \bigintsss_{\Omega_{\text{direct}}} I(\Omega)\, \text{d}\Omega \Big) / \Omega_{\text{direct}} } 
							{\Big( \bigintsss_{\Omega_{\text{rest}}} I(\Omega)\, \text{d}\Omega \Big) / \Omega_{\text{rest}}}
					= \dfrac{I_{\text{direct}}}{I_{\text{rest}}}  \, .
\end{equation}

\section{Directional scattering of a plane wave}

We found that for small populations (\(N_{\text{pop}} \ll 100\)), the optimization does not always converge to the best observed directionality ratio of \(R_{\text{direct}} \approx 12\) (\(10.8\,\)dB).
Small populations seem to offer insufficient genetic diversity and tend to get stuck in local extrema.
The improvement during \(50\) generations of an evolution with a population size of \(N_{\text{pop}} = 500\) is shown in figure~\ref{fig:fig2}(a).
Blue dots correspond to the fitness of the individuals, the green and red lines indicate the best and average fitness, respectively.
The best solutions during successive iterations and their far-field backscattering patterns are shown in Fig.~\ref{fig:fig2}(b)-\ref{fig:fig2}(g) and Fig.~\ref{fig:fig2}(h) shows the best solution after the \(50^{\text{th}}\) generation.
To verify the convergence and reproducibility, we repeated the same optimization with random initial populations, yielding always similar antenna morphologies with directionality ratios \(R_{\text{direct}} \approx 10.8\,\text{dB}\) (see Appendix, Fig.~\ref{figApp:reproducibility}).

At first sight we observe that the optimized nano-antenna visually resembles a Yagi-Uda RF antenna \cite{yagi_beam_1928}.
An analysis of the phase of the electric field \(E_x\) component reveals a difference of \(\pi\) between the central and the outer parts of the nanostructure (Fig.~\ref{fig:fig2}(i)).
We identify three main functional elements, as shown in Fig.~\ref{fig:fig2}(j):
A driving feed in the antenna center (blue), a reflector on the left (green) and a director on the right (red).
\added{As we explain in more detail in the appendix (see also Fig.~\ref{figApp:functional_elements_analysis}), we want to note that this is not a categorical analogy to RF antenna design, but rather a simplified picture giving an idea about the roles of the different parts of the structure.}

\added{The intrinsic phase differences from resonance de-tuning of the elements in combination with the adequate distances between reflector, feed and director, result in cancellation of the scattered fields in backward direction and in constructive interference in forward direction.}
In fact, this working principle is \added{equivalent} to the RF Yagi-Uda antenna, \added{except that in our case the scattering is occurring at an angle with respect to the antenna axis}.
Actually it has been shown earlier, that RF antenna concepts can work also at optical frequencies, despite large plasmonic losses \cite{hofmann_design_2007, novotny_effective_2007}.

Nano Yagi-Uda antennas have been demonstrated at several occasions \cite{curto_unidirectional_2010, ramezani_hybrid_2015, xiong_compact_2016}.
Through an analysis of the impact of a removal or insertion of a gold block at every possible position (Fig.~\ref{fig:fig2}(k)) we find that the possible further increase in directionality is small. Furthermore we observe that fusing single blocks in the reflector and director, or cutting through the feed element kills the directionality. 
This is because changing the length of an element would result in a spectral shift of its LSP resonance, which would break the correct relative phases and scattered field amplitudes required for directional scattering.

\begin{figure}[tb]
	\centering
	\includegraphics{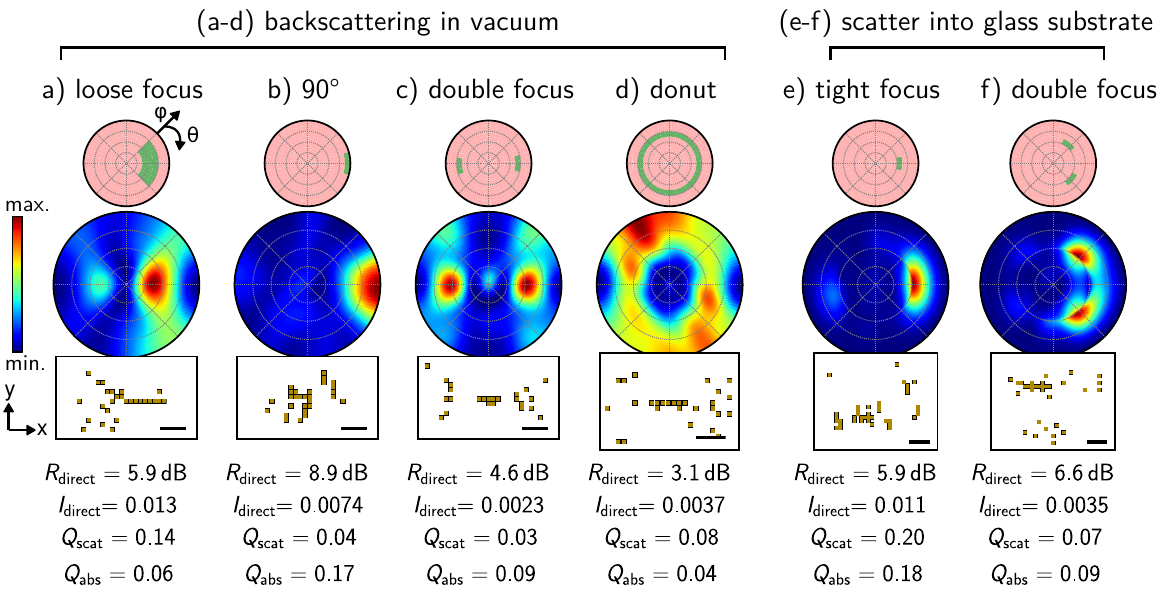}
	\caption{\FIGCAPTIONPREFIX
		Examples of directional EO (maximize \(R_{\text{direct}}\)) for various target solid angles (green areas in top plots). Bottom plots show top views of the optimized geometries, where scale bars are \(200\,\)nm. 
		Center plots show scattering patterns for the full hemisphere of a normally incident plane wave at \(\lambda_0 = 800\,\)nm, linearly \(X\)-polarized. 
		(a) towards a loose focal area, (b) towards \(90^{\circ}\), (c) bidirectionally and (d) towards a ``donut'' target area.
		(a-d) are set in a vacuum environment and show backscattering. 
		(e-f) are set on an \(n_{\text{s}}=1.5\,\) dielectric, hemispherical substrate in order to reflect the conditions in oil immersion microscopy. In these two cases, forward scattering into the glass substrate is maximized for areas at the critical angle.
	}
	\label{fig:fig3}
\end{figure}

\subsection{Different directional scattering objectives}

In a second step, we run several optimizations for directional scattering problems, shown in figure~\ref{fig:fig3}.
The top row polar plots indicate the target solid angles \(\Omega_{\text{direct}}\) by green color, which are of varying size and position.
The bottom plots show top views of the optimized gold block arrangements.
The center plots show the scattering radiation patterns of the optimization outcome.
In order to demonstrate that directionality is not obtained mainly through the index contrast at the substrate, we also show optimizations with different target solid angles for a vacuum environment (\(n_{\text{env}}=1\), Figs.~\ref{fig:fig3}(a)-\ref{fig:fig3}(d)).
Figs.~\ref{fig:fig2}, \ref{fig:fig3}(e)~and~\ref{fig:fig3}(f) show optimizations for plasmonic structures on a dielectric substrate (\(n_{\text{s}}=1.5\)), where in \ref{fig:fig3}(e)~and~\ref{fig:fig3}(f) furthermore scattering into the substrate was maximized. In all other cases scattering goes towards the air hemisphere.

While EO yields good directionality ratios in all cases, we observe that the scattering intensity and scattering efficiency \(Q_{\text{scat}}\) (ratio of scattering and geometric cross sections) are generally very low. 
The highest scattering intensities are obtained for large target solid angles and small deflection angles (Fig.~\ref{fig:fig3}(a)) or for scattering into a substrate at the critical angle (Fig.~\ref{fig:fig3}(e)). But even those cases have very small scattering efficiencies well below~\(1\).
As a first conclusion we record that it is a physically demanding task to steer an incident plane wave unidirectionally at large angles via a planar gold nanostructure.

\begin{figure}[tb]
	\centering
	\includegraphics{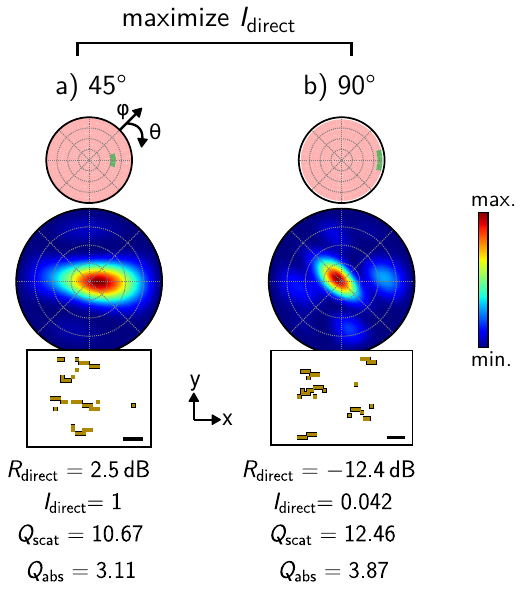}
	\caption{\FIGCAPTIONPREFIX
		EO for maximization of the absolute backscattered intensity \(I_{\text{direct}}\) to the target solid angle centered at a polar angle of (a) \(45^{\circ}\) and (b) \(90^{\circ}\).
		Target solid angles, backscattering radiation patterns and top views of the optimized geometries are shown respectively from top to bottom. Scale bars are \(200\,\)nm. 
		Normally incident plane wave, linearly \(X\)-polarized with \(\lambda_0 = 800\,\)nm, the structures lie in vacuum on an \(n=1.5\) substrate. 
		A visualization of the optimization convergence for case (a) can be found online ({\color{blue}Visualization~1}).
	}
	\label{fig:fig3-2}
\end{figure}

\added{For comparison we ran EOs maximizing the absolute backscattered intensity \(I_{\text{direct}}\) towards polar angles of \(45^{\circ}\) and \(90^{\circ}\), shown in Figs.~\ref{fig:fig3-2}(a) and~\ref{fig:fig3-2}(b), respectively.
In both cases, the largest absolute scattered intensity towards the target solid angle is found for a structure built of several resonant gold elements, giving 2-3 orders of magnitude higher overall scattering efficiencies. 
But most of the light is backscattered with no deflection, directionality is completely lost.
Still, the scattered intensity \(I_{\text{direct}}\) towards the target solid angles in Fig.~\ref{fig:fig3-2} is several times higher compared to Fig.~\ref{fig:fig2} and Fig.~\ref{fig:fig3}(b), where the ratio \(R_{\text{direct}}\) was maximized. The latter antennas are built from detuned elements, resulting in a very weak absolute scattering.}
This supports our conclusion that using planar plasmonic nanostructures, it is physically almost impossible to redirect a normally incident plane wave towards the direction of its linear polarization. 
\added{Large scattering intensities seem to imply that almost no directionality is obtained.}

\added{We note that Fig.~\ref{fig:fig3-2} shows structures on a substrate. We also performed the same optimization in a vacuum environment, yielding similar results.}

We provide a visualization online ({\color{blue}Visualization 1}), showing the progress of EO of a large-area (\(5 \times 5\,\)\textmu m\(^2\)), intensity optimized scatterer (analog to Fig.~\ref{fig:fig3-2}(a)). It results in an assembly of several resonant nanoantennas, arranged in a grating-like way. Despite the large allowed area and consequently much more degrees of freedom, also this optimization fails to combine a good directionality ratio with high scattering intensity.

\subsection{Discussion: Weak scattering due to phase difference requirement}

The weak scattering at high deflection angles is a direct result of the physical properties of the plane wave illumination. The phase of the illumination field is constant across the entire planar structure, \added{which means that it is not possible to obtain phase differences only through positioning of plasmonic elements}. 
\added{On the other hand, the directional antenna requires a proper phase difference between antenna feed and director elements.}
To achieve this, the EO algorithm choses very small blocks as director elements, which are driven by the plane wave below resonance (\(\omega_{\text{res}} \gg \omega_0\)), resulting in a phase of \(-\pi / 2\) between illumination and response.
As driving feed the optimizer choses a large rod having its LSP resonance in the infrared (\(\omega_{\text{res}} \ll \omega_0\)), which gives a phase of \(+\pi / 2\) with respect to the driving plane wave. 
In consequence, the feed and director structures have a relative phase of \(\pi\) with respect to each other but are also entirely off-resonance and thus scatter only very weakly. This results in an overall small scattering intensity, so most of the incident light passes just through the plasmonic structure without interaction. 
\added{The weak interaction with the incident plane wave is confirmed by the very low absorption efficiencies of the optimized antennas in Figs~2 and~3.}

As we see in Figs.~\ref{fig:fig3-2}(a) and \ref{fig:fig3-2}(b), if the intensity is target of the optimization, the resulting structures look fundamentally different: The dispersed small blocks, that give the directionality-optimized antennas the fragmented look, disappear. The EO algorithm now uses all the available material to construct as many resonant nano-rods as possible, yielding a strong scattering cross section but very weak directionality.

\begin{figure}[t]
	\centering
	\includegraphics[page=1]{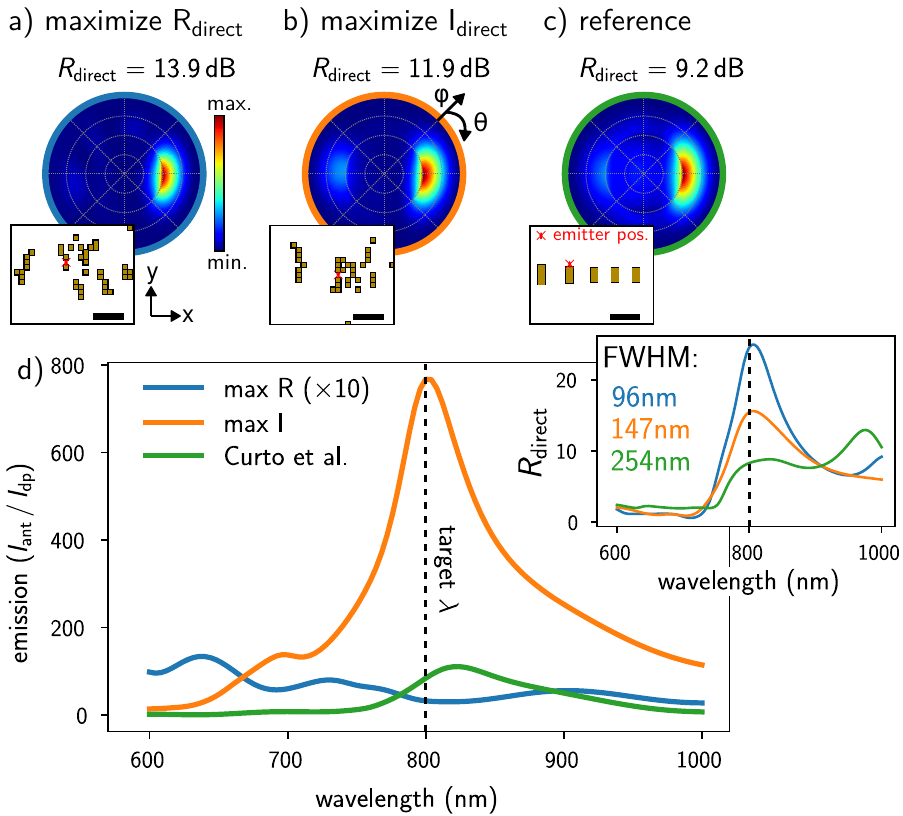} 
	\caption{\FIGCAPTIONPREFIX
		Optimized antennas for dipole emitter directional emission. The quantum emitter is oriented along \(OY\), radiates at \(\lambda_0 = 800\,\)nm and has a minimum distance to the gold structure of \(20\,\)nm. Emission is maximized at the critical angle of scattering inside the \(n_{\text{s}}=1.5\) substrate.
		(a) maximize directionality ratio \(R_{\text{direct}}\).
		(b) maximize absolute intensity \(I_{\text{direct}}\) at target solid angle.
		(c) simulation of reference antenna from Ref.~\onlinecite{curto_unidirectional_2010}.
		(d) spectrum of total dipole emission intensity, normalized to the intensity \(I_{\text{dp}}\) of an isolated dipole emitter (no antenna). 
		The inset shows directionality spectra for the three antennas.
		Radiation patterns are evaluated on a hemispherical screen inside the substrate. Scale bars are \(200\,\)nm.
	}
	\label{fig:fig4}
\end{figure}

\added{We finally want to note, that a drawback of the GDM with regards to fabrication robustness is, that diagonally neighboring metallic blocks are electrically isolated in the simulations. 
In an actual sample however, there are high chances that diagonal neighbor blocks are electrically connected at their touching point due to unavoidable imperfections in the fabrication process. Such electric connections can have a significant impact on localized surface plasmon resonances and in consquence have the potential to change the optical response of the structure.
To assess the robustness of our results against fabricational imperfections, we ran simulations, connecting diagonally touching elements in the optimized antennas using additional dipoles, which didn't have a severe impact in these cases.}
To generally overcome this shortcoming, dielectric / plasmonic hybrid structures may be designed which can yield inherent phase differences without requiring precise tailoring of the relative LSP resonances \cite{zhang_unidirectional_2019}.

\section{Quantum emitter directional antenna}

As found above, plane wave directional scattering at high scattering efficiency is physically difficult to obtain. 
On the other hand, it has been demonstrated that plasmonic nanostructures can be used as efficient directional antennas for quantum emitters, being also able to boost the decay of a quantum transition thanks to the Purcell effect \cite{curto_unidirectional_2010, maksymov_optical_2012, vercruysse_directional_2014, hancu_multipolar_2014, francs_plasmonic_2016}.
In fact, in the case of a local dipole transition as light source, there is not the problem that the illumination excites every part of the antenna simultaneously and with constant phase. This makes it significantly easier to find a solution.

Following the configuration reported in~\onlinecite{curto_unidirectional_2010}, we replace the plane wave illumination by a dipole transition along \(OY\), oscillating at \(\lambda_0 = 800\,\)nm, positioned at the origin of the coordinate system. 
We prohibit the EO algorithm to put a gold block at this location, which results in a minimum distance of \(20\,\)nm between gold and emitter. The emission is to be maximized towards \(OX\) along the critical angle into the \(n_{\text{s}}=1.5\) substrate.
The EO converges towards an antenna with a directionality ratio as high as \(R_{\text{direct}} \approx 13.9\,\)dB, shown in Fig.~\ref{fig:fig4}(a).
Ref.~\onlinecite{curto_unidirectional_2010} uses a slightly different definition of directionality, therefore we simulate the antenna ``YU145'' of Ref.~\onlinecite{curto_unidirectional_2010} with the GDM, using the same emitter-feed distance of \(20\,\)nm as for the EO. We find a directionality of \(R_{\text{direct, ref.}} \approx 9.2\,\)dB (Fig.~\ref{fig:fig4}(c)).
While this is lower compared to the optimized antenna, we see in Fig.~\ref{fig:fig4}(d) that the emission intensity of the optimized antenna is around two orders of magnitude weaker.
We re-run the EO using as optimization target the absolute intensity of directed emission (Fig.~\ref{fig:fig4}(b)).
While directionality is moderately reduced to \(11.9\,\)dB, the signal from the intensity-optimized directional nanoantenna (orange line in Fig.~\ref{fig:fig4}(d)) is now almost one order of magnitude higher than that of the reference. 
\added{Having a closer look at the geometry in Fig.~\ref{fig:fig4}(b), we observe that the EO algorithm designed the antenna such that the emitter is surrounded by gold to maximize the Purcell effect, which strongly boosts the emitted intensity.}
Finally we find that an increased directionality comes at the cost of a reduced spectral full width at half maximum (FWHM) of \(R_{\text{direct}}\), hence leads to a narrower operation window (see inset Fig.~\ref{fig:fig4}(d)).

\begin{figure}[t]
	\centering
	\includegraphics[width=\textwidth]{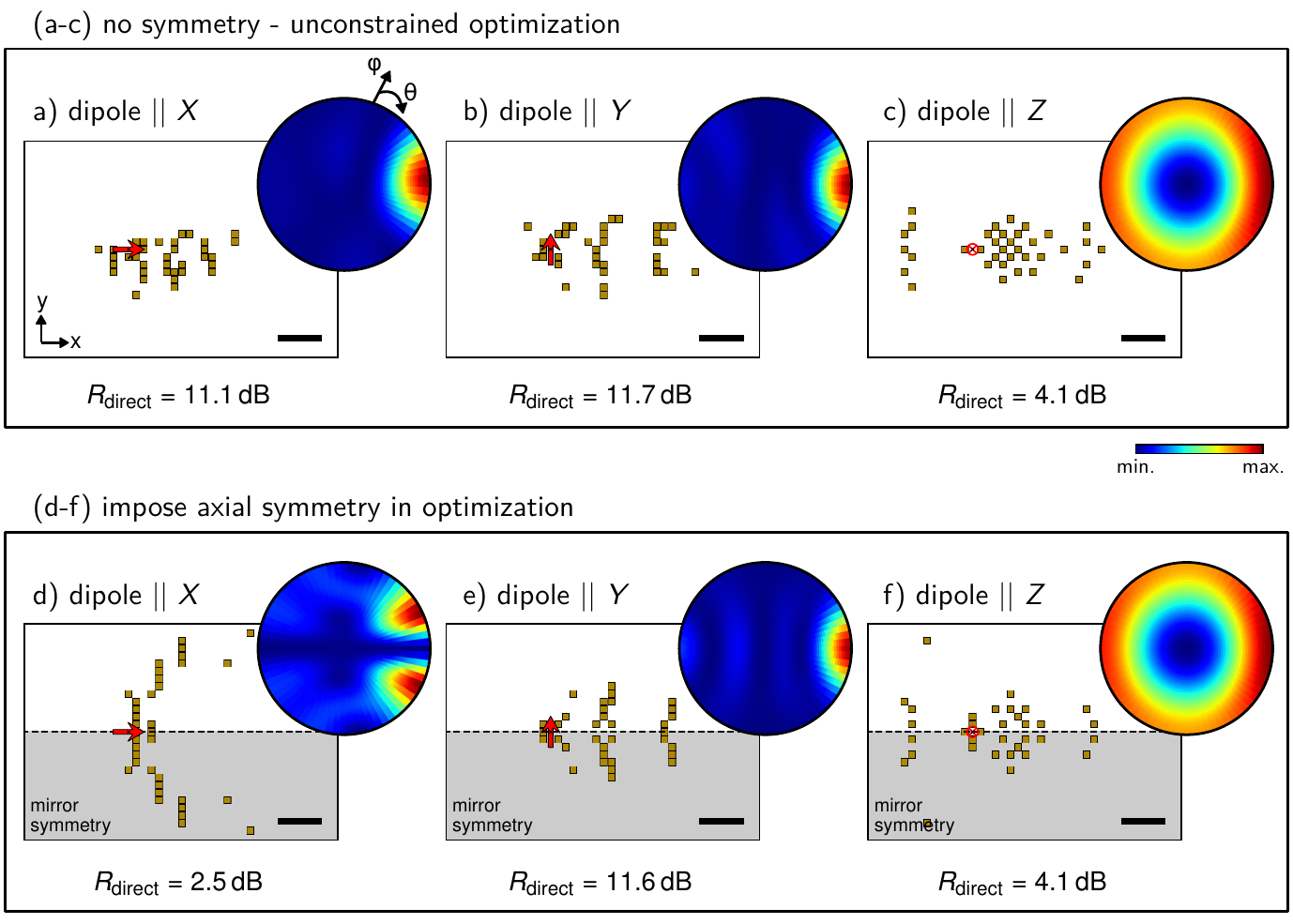}
	\caption{\FIGCAPTIONPREFIX
		Symmetry analysis, showing optimization results of directional quantum emitter gold antennas in vacuum environment (\(n{\text{env}}=1\)) for different orientations of the dipole transition. 
		(a) and (d): dipole emitter along \(OX\), (b) and (e): dipole emitter along \(OY\), (c) and (f): dipole emitter along \(OZ\).
		In (a-c), the geometry is unconstrained, in (d-f), the same optimizations were repeated, imposing an axial mirror symmetry at the \(x\)-axis for the geometry of the antenna.
		Emitter placed at \((x=0\,\text{nm}, y=0\,\text{nm}, z=20\,\text{nm})\), indicated by a red arrow. 
		Scale bars are \(200\,\)nm.
	}
	\label{fig:fig5}
\end{figure}

\subsection{Dipole emitter orientation}

To study the role of the orientation of the quantum emitter's dipole transition, we run antenna optimizations for \(X\), \(Y\) and \(Z\) oriented dipoles, shown in Figs.~\ref{fig:fig5}(a)-\ref{fig:fig5}(c).
To keep the symmetry of the system as high as possible, we assume a vacuum environment and maximize scattering towards \(OX\) (solid angle as used in Fig.~\ref{fig:fig3}(b)).
In case of \(X\)- and \(Y\)-oriented dipoles (Figs.~\ref{fig:fig5}(a)-\ref{fig:fig5}(b)), the optimization organizes the gold blocks in a compact arrangement, forming a strongly directional antenna.
On the other hand, for a dipole pointing out of the structure plane (\(Z\) direction), the limited height of the planar gold structure (\(H=40\,\)nm) impedes obtaining a significant optical response of the gold blocks at the emission wavelength of \(\lambda_0=800\,\)nm, hence directionality remains weak (Fig.~\ref{fig:fig5}(c)).

\subsection{Antenna symmetry}

In contrast to the intuitive expectation that symmetric antennas should perform best, EO finds mostly rather fragmented antennas of low symmetry.
To assess the role of symmetry, we re-run the quantum emitter antenna optimizations for different dipole-orientation (Figs.~\ref{fig:fig5}(a)-\ref{fig:fig5}(c)), imposing a mirror symmetry around the \(X\) axis.
The results shown in Fig.~\ref{fig:fig5}(d)-\ref{fig:fig5}(f) suggest, that symmetric antennas give a similar performance in the cases of dipole transitions aligned along \(Y\) and \(Z\), as depicted in Fig.~\ref{fig:fig5}(e), respectively~(f), even though the symmetric designs do not yield noticeable improvement over an unconstrained geometric model.

In the case of an \(X\)-oriented dipole however, the symmetric EO completely fails to design a directional antenna (Fig.~\ref{fig:fig5}(d)).
By inspecting the antenna in Fig.~\ref{fig:fig5}(a), we notice that it is designed such that the polarization of the \(X\)-oriented dipole emitter is converted by a gold element to a plasmonic current along \(Y\), effectively generating an accordingly oriented plasmonic dipole emitter.
The scattered light of this LSP dipolar resonance is then guided towards the target solid angle via a Yagi-Uda-type assembly of structures around the feed.
By imposing an axial symmetry at the dipole position, this local polarization conversion is physically impossible. 
Instead, in Fig.~\ref{fig:fig5}(d) the EO algorithm designs a quadrupolar feed element, directing the light close to the desired solid angle. However, due to symmetry, emission is zero towards \(X\).

\section{Conclusions}

In conclusion, we demonstrated how evolutionary optimization can be used to design directional plasmonic nanoantennas.
In our approach, within the limits of possible geometries built with \(40\) equal gold blocks, the design of the structures is completely free.
We showed, that EO automatically finds a nanostructure layout which corresponds to well known design-principles from radio-frequency antennas.
By comparing the performance of directionality- and intensity-optimized planar gold nanostructures, we found that high directionality ratios for scattering of normal incidence plane waves to large angles comes necessarily at the cost of weak scattering intensities.
We demonstrated that in contrast to weak directional plane wave scattering, EO can design very efficient directional nanoantennas for quantum emitters. 
Via an appropriate fitness function, EO finds an antenna which not only yields very good directionality but also exploits at best the Purcell effect for quantum emitter signal enhancement. 
Through an analysis of the EO results for different fitness functions we furthermore found, that high directionality ratios come at the cost of spectrally less broadband working windows.
Finally, through an EO-based analysis of symmetry conditions we found that, dependent on the orientation of the quantum emitter dipole transition, it can be necessary to design and asymmetric nano-antenna for good performance.

We conclude, that from the results of EO general design rules can be derived for the tailoring of photonic nanostructures. 
EO can also be used as a benchmark tool in nano-photonic design and can reveal intrinsic physical limitations of nano-optical configurations.
We foresee that EO has manifold further promising applications beyond directional scattering, for examples in field enhanced spectroscopy or non-linear nanophotonics.

\section*{Supplementary materials}

Visualization 1: Movie showing the convergence process of EO for an intensity-optimized, plane wave illuminated, directional optical antenna.

\section*{Appendix}

\subsection*{Reproducibility}
\begin{figure}[h]
	\centering
	\includegraphics[width=\textwidth,page=1]{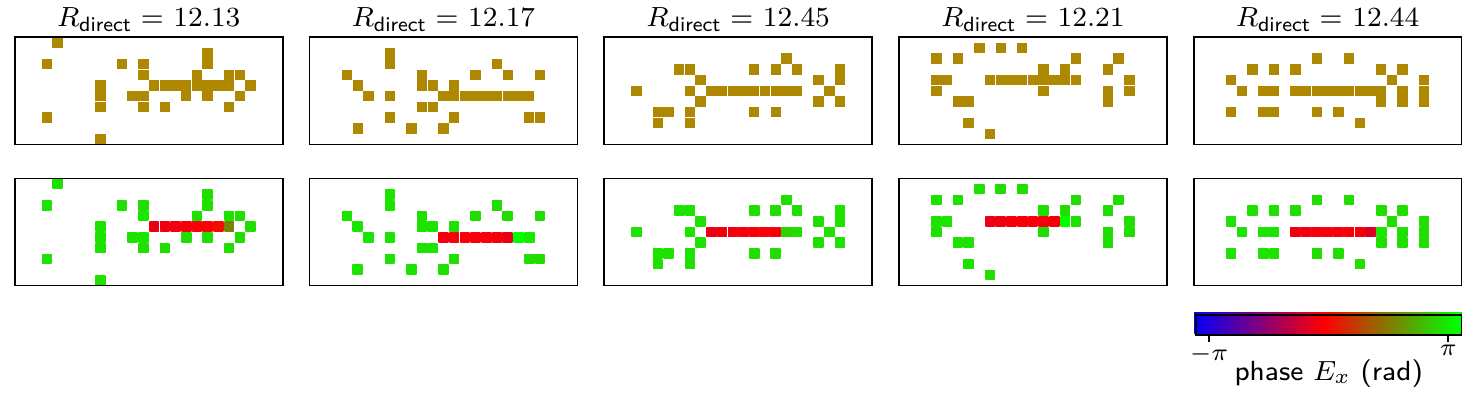} 
	\caption{\FIGCAPTIONPREFIX
		Demonstration of reproducibility and convergence: Top row shows antenna geometries and their directivity ratios for \(5\) independent runs of the same optimization, with random initial EO populations.
		Bottom row shows the relative phase of the \(E_x\) field component.
	}
	\label{figApp:reproducibility}
\end{figure}

To test the convergence and the reproducibility of the EO, we re-run the optimization shown in main text Fig.~\ref{fig:fig2} several times with different, randomly initialized starting populations.
As shown in Fig.~\ref{figApp:reproducibility}, the final antennas converge always to the same maximum directionality ratio of \(R_{\text{direct}} \approx 12\).
The geometries of the solutions are in all cases similar, leading to the same phase distribution.

\subsection*{Functional elements of plane wave directional antenna}

\begin{figure}[h]
	\centering
	\includegraphics[width=\textwidth,page=1]{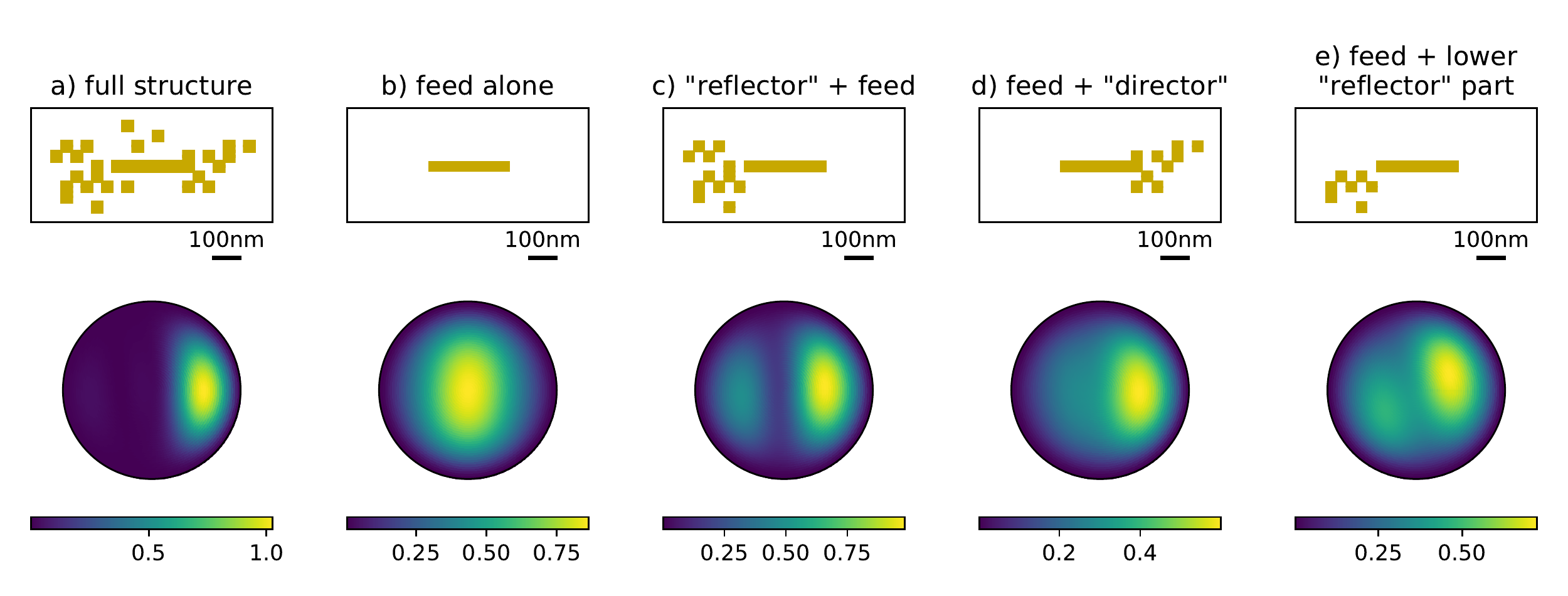} 
	\caption{\FIGCAPTIONPREFIX
		Analysis of functional elements for the antenna in Fig.~\ref{fig:fig2}. Top row: Top view of the considered sub-part of the full antenna geometry. 
		Bottom row: backscattering pattern.
		Scale bars are \(100\,\)nm, the colorscale represents the scattering intensity in arbitrary units (normalized to the peak intensity of the full structure).
	}
	\label{figApp:functional_elements_analysis}
\end{figure}

It is clear that the parasitic elements around the feed lead to the directional scattering effect in the plane-wave antennas (as demonstrated in Fig.~\ref{figApp:functional_elements_analysis}(a) and~\ref{figApp:functional_elements_analysis}(b) below). 
On the other hand it is difficult to ascribe to specific clusters of parasitic elements the unambiguous role of a ``reflector'' or a ``director''.
To assess whether it is justified to call the agglomeration of blocks on the left of the feed in Fig.~\ref{fig:fig2} an equivalent of a ``reflector'' and the agglomeration of blocks on the right a ``director'', we simulated the feed alone, the feed + ``reflector'' and the feed + ``director'', shown in Fig.~\ref{figApp:functional_elements_analysis}(b)-\ref{figApp:functional_elements_analysis}(d).

While we observe no directional scattering for the isolated feed (Fig.~\ref{figApp:functional_elements_analysis}(b)), the reflector as well as director alone suffice to impose a directional effect on the feed element (Figs.~\ref{figApp:functional_elements_analysis}(c)-\ref{figApp:functional_elements_analysis}(d)), in analogy to what would be expected in an RF Yagi-Uda antenna.
On the other hand, if only a part of for example the reflector cluster is used, the direction of scattering becomes tilted with respect to the target angle (Fig.~\ref{figApp:functional_elements_analysis}(e)). The full ensemble of ``reflector'' blocks is necessary to obtain the desired scattering effect.
We conclude that it is adequate to consider the left and right agglomerations of parasitic elements equivalent to a ``reflector'' and a ``director''. 

As a final note, we want to stress that this is not a categorical analogy. 
A direct adaptation of the structure to RF antennas is not easily possible, we rather consider our analogy a simplified description of the role of the complex arrangements of parasitic elements.

\subsection*{Discretization, size and shape of elementary blocks}

In principle any kind of constituents can be used with the EO-GDM method to compose the structure.
However, larger building blocks or elementary blocks of some complex shape constrain the generality of possible geometries. 
On the other hand, small blocks allow to perform a more complete search of the solution space, but at the cost of slower convergence.
The choice of size (and shape) of the elementary blocks is therefore always a compromise between how flexible the geometric model should be and the speed of convergence of the optimization.
If constraints to the attainable diversity of geometries are acceptable, only few but larger building blocks and/or blocks of more specific shape like cuboids, rods or polygonal structures might be chosen as elementary unit.
In view of a possible experimental realization of the structures, it makes also sense to use building blocks which are not smaller then the structural resolution that can be obtained in fabrication.

\begin{figure}[h]
	\centering
	\includegraphics[width=\textwidth,page=1]{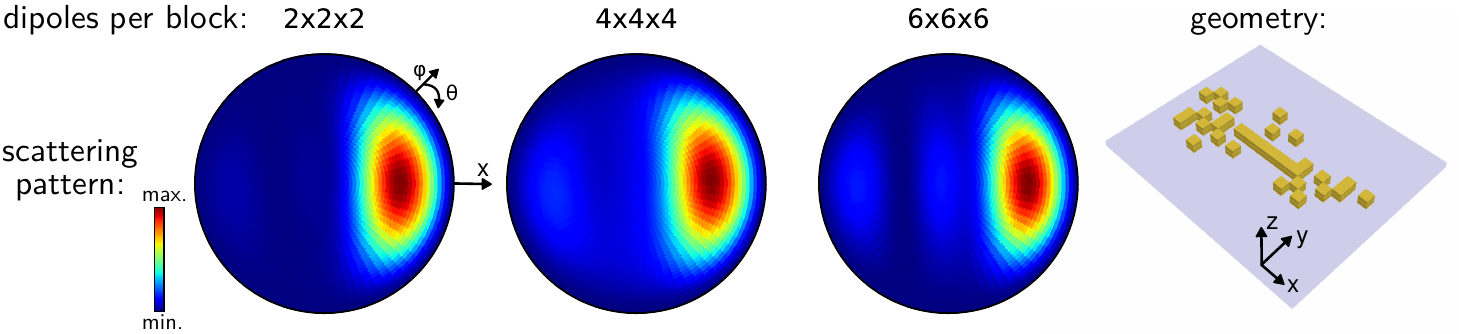} 
	\caption{\FIGCAPTIONPREFIX
		Backscattering patterns of an identical gold nanostructure but using different discretization steps (geometry shown on the right, same as in Fig.~\ref{fig:fig2}). 
		\(\lambda=800\,\)nm plane wave illumination from the top, linear polarized along \(X\).
		Each elementary block in the structure was discretized from left to right by \(2\times 2\times 2\) dipoles with \(20\,\)nm step, \(4\times 4\times 4\) dipoles with \(10\,\)nm step and \(6\times 6\times 6\) dipoles with \(6.67\,\)nm step.
		The results are very similar, hence we can conclude that a step of \(20\,\)nm is a good approximation.
	}
	\label{figApp:discretization_step}
\end{figure}

Concerning the discretization, we used steps of \(20\,\)nm which gives good optimization speed because the full structure consists of only \(40\times (2\times 2\times 2) = 320\) dipoles.
In order to verify that this rather coarse discretization is a good approximation, we re-calculated the structure shown in figure~2 with finer step-sizes.
As shown in Fig.~\ref{figApp:discretization_step}, simulations with steps of \(20\,\)nm (\(2\times 2\times 2\) dipoles per block), \(10\,\)nm  (\(4\times 4\times 4\) dipoles per block) and \(6.67\,\)nm (\(6\times 6\times 6\) dipoles per block) yield very similar results, so our approximation using steps of \(20\,\)nm is good.



\section*{Funding}
This work was supported by Programme Investissements d'Avenir under the program ANR-11-IDEX-0002-02, reference ANR-10-LABX-0037-NEXT and by the computing facility center CALMIP of the University of Toulouse under grant P12167.
PW received funding by the German Research Foundation (DFG) through a Research Fellowship (WI 5261/1-1). OM acknowledges support through EPSRC grant EP/M009122/1.

\section*{Acknowledgments}
We gratefully thank N.~Bonod for fruitful discussions.
All data supporting this study are openly available from the University of Southampton repository (DOI: 10.5258/SOTON/D0992).








\end{document}